\documentclass[a4paper]{iopart}
\usepackage{epsfig}
\usepackage{dcolumn}
\usepackage{epstopdf}
\usepackage[T1]{fontenc}
\usepackage[latin9]{inputenc}
\usepackage{amstext}
\usepackage{graphicx}

\begin{document}

%\title[Thermoelectricity in Mg$_3$Sb$_2$]{Thermoelectricity in Mg$_3$Sb$_2$ vs temperature and doping: an energy- and temperature-dependent relaxation-time theoretical assessment}
\title[Theory of thermoelectricity in Mg$_3$Sb$_2$]{Theory of thermoelectricity in Mg$_3$Sb$_2$ with an energy- and temperature-dependent relaxation time}
\author{Roberta Farris, Maria Barbara Maccioni, Alessio Filippetti, {\rm and} Vincenzo Fiorentini}

\address{Department of  Physics at University of Cagliari, and CNR-IOM, UOS Cagliari, Cittadella Universitaria, I-09042 Monserrato (CA), Italy}
%\ead{roberta.farris@dsf.unica.it, vincenzo.fiorentini@gmail.com}

\begin{abstract}
We study the electronic transport coefficients  and the thermoelectric figure of merit ZT in $n$-doped Mg$_3$Sb$_2$  based on density-functional electronic structure and Bloch-Boltzmann transport theory with an energy- and temperature-dependent relaxation time. Both the  lattice and electronic thermal conductivities affect the final ZT significantly, hence we include the lattice thermal conductivity calculated ab initio.
Where applicable, our results are in good agreement with existing experiments, thanks to the treatment of lattice thermal conductivity and the improved description of electronic scattering.  ZT increases monotonically in our T range (300 to 700 K), reaching a value of 1.6 at 700 K; it  peaks as a function of doping at about 3$\times$10$^{19}$ cm$^{-3}$. At this doping, ZT$>$1 for T$>$500 K.  \end{abstract}
%\maketitle
\section{Introduction}
Thermoelectrics convert heat  into electrical power, or thermal gradients into electrical potential gradients, thus making possible the  recycling of thermal waste into usable energy. A measure of the quality of a thermoelectric material is the figure of merit 
\begin{equation}
 {\rm ZT}=\frac{\kappa'}{\kappa}-1,
\end{equation}
where $\kappa'$ and $\kappa$ are the thermal conductivity, respectively, in the presence  and in the absence of an electric field. ZT  essentially quantifies the strength of the component of the heat flow set by the (thermo)electric field relative to the total heat flow \cite{fund}. In the absence of electric current, it is easy to show from the constitutive thermoelectric relations that 
\begin{equation}
{\rm ZT}=\frac{\ \sigma {\rm S}^2}{\kappa}{\rm T}=\frac{\sigma {\rm S}^2}{\kappa_e +\kappa_{\ell}}{\rm T},
\label{ZTdef}
\end{equation}
where $\sigma$ is the electrical conductivity, $\kappa$, ${\kappa_e}$ and $\kappa_{\ell}$  the total, electronic, and lattice thermal conductivities, S the Seebeck coefficient, and T the temperature.  Mg$_3$Sb$_2$ and related compounds  have emerged as interesting candidates \cite{mgsb1,gb}, and are the test case for this paper. 

Maximizing ZT is a complicated business. A large parameter space needs to be explored: intrinsic (structure and thermal properties, band valleys, effective masses)  and extrinsic (doping, temperature) to the material, as well as conceptual (scattering mechanisms, electronic structure approximations). The various ingredients often combine and intertwine fairly perversely; for example the electrical conductivity appears in the numerator, but also hides in the thermal conductivity in the denominator ($\kappa_e$ and $\sigma$ are closely related by the Lorenz formula, as well as more in general by Eq.\ref{coeffic} below) as well as, even more subtly, in the Seebeck coefficient (by the Mott-Cutler formula \cite{mott}, which also derives from Eq.\ref{coeffic}). The lattice thermal conductivity is instead largely decoupled from doping and from electrical transport. 

As a general guideline, a low thermal conductivity is ostensibly a bonus: luckily, it is know experimentally \cite{mgsb1} that the lattice component in Mg$_3$Sb$_2$-based materials can be as low as 1 W/(K m); we have recently theoretically clarified  \cite{kappa} that this is due to polycrystallinity in Mg$_3$Sb$_2$. Of course, a large power factor $\sigma$S$^2$ and a low electronic thermal conductivity are also needed, and in this paper we will focus mainly on these coefficients of electronic origin, making sense of how they all conspire to eventually produce a good ZT figure of merit. 

While a fully ab initio study of the thermoelectric  coefficients is  impossible, much headway can be made describing the electronic band structure from first principles, and then using the band structure in  a linearized Boltzmann transport equation for electrons in the relaxation-time approximation, in a version known as Bloch-Boltzmann theory \cite{allen}, using an energy- and temperature-dependent relaxation time. Further progress can be achieved in principle using ab initio electron-phonon scattering rates \cite{fb,giu,vdw,ponce}, but this approach is considerably more complex and is still in its infancy in the field of thermoelectricity. 

A popular implementation of this transport theory is in the BoltzTrap \cite{bt} code,    assuming rigid  (i.e. not changing with doping or temperature) bands and a constant relaxation time $\tau_0$. BoltzTrap computes the electronic transport coefficients relevant to thermoelectricity, namely the electrical conductivity $\sigma$, the electronic thermal conductivity $\kappa_e$, and the Seebeck thermopower coefficient S, as
\begin{equation}
\sigma={\cal L}^{(0)}; \ \ 
{\rm S}=\frac{{\cal L}^{(1)}}{e {\rm T} {\cal L}^{(0)}};\ \  
\kappa_e=\frac{1}{e^2{\rm T}}\left(\frac{({\cal L}^{(1)})^2}{{\cal L}^{(0)}} -{\cal L}^{(2)}\right)
\label{coeffic}
\end{equation}
with $e$ the electronic charge, and 
\begin{equation} \label{eqL}
    \mathcal{L}^{(N)}=e^{2}\int K(E,T)(E-\mu)^{N}\left(-\frac{\partial f_F(E;\mu,T)}{\partial E}\right)dE.
\end{equation}Here
\begin{equation} \label{eq:2.2}
    K(E,T)=\int \sum_{b} \,\mathbf{v}_{b,\mathbf{k}}\otimes \mathbf{v}_{b,\mathbf{k}}\, \tau_{b,\mathbf{k}}\, \delta(E-E_{b,\mathbf{k}})\frac{d\mathbf{k}}{8\pi^{3}}
\end{equation}
where $b$ is a band index, {\bf k} is the conserved wavevector, $E_{b,\bf k}$ the band energies, $\mathbf{v}_{b,\mathbf{k}}$ the corresponding velocities,  $f_F$ the Fermi distribution, $\mu$ the chemical potential, and $\tau$ the  relaxation time (generally a function of {\bf k}, $E$, and T).
It also  produces smooth {\bf k}-dependent bands on very fine grids by interpolating (with a Wannier-like expansion) the energies computed ab initio on comparatively coarser grids.

\begin{figure}[ht]
\centerline{\includegraphics[width=0.6\linewidth]{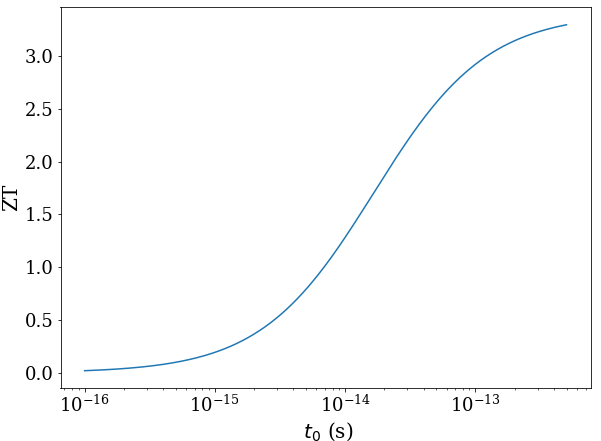}}
\caption{\label{ZT0} ZT vs $\tau_0$ in the constant time approximation (for Mg$_3$Sb$_2$ calculated parameters at 500 K and 3$\times$10$^{19}$ cm$^{-3}$ $n$-doping.)}
\end{figure}

By way of  motivation for exploring improved approximations beyond  the 
  constant relaxation time approximation, we note that  a constant $\tau_0$ factorizes out of Eq.\ref{eqL}, hence the code returns just the quantities $\sigma_0$=$\sigma$/$\tau_0$ and $\kappa_{e,0}$=$\kappa_e$/$\tau_0$, which are determined  by the band-structure and by temperature (via the occupation function) but are independent of $\tau_0$ (the constant relaxation time cancels out of the Seebeck). If the lattice thermal conductivity  $\kappa_{\ell}$ were zero, $\tau_0$ would cancel out of ZT, which would then be  a T-independent constant. But since $\kappa_{\ell}$ is not zero, ZT (Eq.\ref{ZTdef}) becomes in this approximation $${\rm ZT}=\frac{\sigma_0\tau_0 S^2}{\kappa_{e,0}\tau_0+\kappa_{\ell}},$$
which depends crucially on $\tau_0$  precisely in the region of typical expected values, as shown in Fig.\ref{ZT0}. Indeed \cite{cardona} even an energy-averaged $\tau$ is expected to be significantly temperature-dependent (typically via a power law)  due  both to  energy averaging and to explicit energy dependence in $\tau_{b,\mathbf{k}}$, e.g. via phonon occupations. Thus, a direct estimate of the dependence of $\tau$ on  energy and temperature seems advisable,  even though on a phenomenological basis.  
(Further refinements beyond our present scope could account for the {\bf k} dependence of $\tau$, in particular as regards electron-phonon scattering: this was the basis for the explanation of the positive thermopower in metallic Li \cite{Li}.)

All this considered, in this paper we go beyond  the zero-order, popular \cite{mgsb1} but basic, constant-$\tau$ approximation and employ analytical energy-dependent  expressions for the relaxation time, which were developed on the basis of known
semiclassical theories \cite{boo} and include the most important mechanisms of electron scattering by acoustic phonons, polar-optical phonons, and charged impurities.  The scattering rates of these mechanisms (further details can be found in \cite{boo,Ridley,thermoCA}) are summarized below; they have been
implemented by our group in BoltzTrap, and usefully employed  for a number of different
systems with satisfactory results \cite{thermoCA}. 
Acoustic phonon scattering is treated within the elastic deformation potential approach in the  long-wavelength acoustic-phonon limit, and the scattering rate is
\begin{equation} 
P_{\rm ac}\left({\rm T},E\right )=
\frac{\left(2\overline{m}\right)^{\frac{3}{2}}  k_{B}{\rm T} D^2 \sqrt{E}}{2\pi\hbar^4\rho v^2}
\label{tau_analitic}
\end{equation} 
where $E$is the electron energy, $D$ the deformation potential of the band energies calculated at the band extrema, $\rho$ the mass density, $v$ the average sound velocity, $\overline{m}$ the  average effective mass.
 Polar optical scattering is modeled following Ridley \cite{Ridley}. The total rate is \begin{equation}
P_{\rm polar} ({\rm T},E)=\sum_i\, 
\frac{C({\rm T}, E, e_i^{\rm LO}) 
-A({\rm T}, E, e_i^{\rm LO}) 
-B({\rm T}, E, e_i^{\rm LO}) }{\ Z({\rm T}, E, e_i^{\rm LO}) \,  E^{3/2}}
\label{tau_drift}
\end{equation} 
where the sum is over all longitudinal-optical phonons, with energy $e_{i}^{\rm LO}$; 
the functions  $A$, $B$, $C$, and $Z$ are omitted for brevity and can be found in Ref.\cite{Ridley}. 
For impurity scattering we adopt the  Brooks-Herring formula
\begin{equation}
P_{\rm imp}\left({\rm T},E\right)=
\frac{\pi n_I Z^2_Ie^4 E^{-3/2}}{\sqrt{2\overline{m}}\left(4 \pi \epsilon_0 \epsilon\right)^2} 
\left[\log{\left(1+\frac{1}{x}\right)}
-\frac{1}{1+x}\right]
\label{Brooks-Herring}
\end{equation}
with $$
x=\frac{\hbar^2q^2_0}{8\overline{m} E},
$$
$n_I$ the ionized impurity concentration, $Z_I$ the impurity charge (we assume $Z_I$=1 here), $\varepsilon_0$ and $\varepsilon$ the vacuum permittivity and the relative dielectric constant, and $q_0$=$\sqrt{e^2n_I/(\varepsilon_0\varepsilon k_BT)}$ the Debye screening wavevector. The  global relaxation time is obtained as
$$
\tau({\rm T},E)=\frac{1}{P_{\rm imp}+P_{\rm polar}+P_{\rm ac}}
$$
i.e. the inverse of the total rate. (Piezoelectric scattering is not included since the structure of Mg$_3$Sb$_2$ is non-polar, with space group $R\overline{3}m$.)
These phenomenological expressions require materials parameters, which  have  been calculated via the Quantum Espresso energy and phonon codes \cite{QE}, except for effective masses, that are imported from \cite{mgsb1}. The  values are: high-frequency and total dielectric constants (average) $\varepsilon_{\infty}$=14.2 and $\varepsilon$=26.7; average sound velocity  $v$=2.7 km/s;  LO-phonon  frequencies  $\hbar$$\omega_{\rm LO}$=165, 200 and 250 cm$^{-1}$; conduction-band deformation potential $D$=6.4 eV; average effective mass $m^*$=0.3 $m_e$. When needed for comparison, the constant relaxation time is set to $\tau_0$=1.34$\times$10$^{-14}$ s, also from Ref.\cite{mgsb1}.

\begin{figure}[ht]
\centerline{\includegraphics[width=0.5\linewidth]{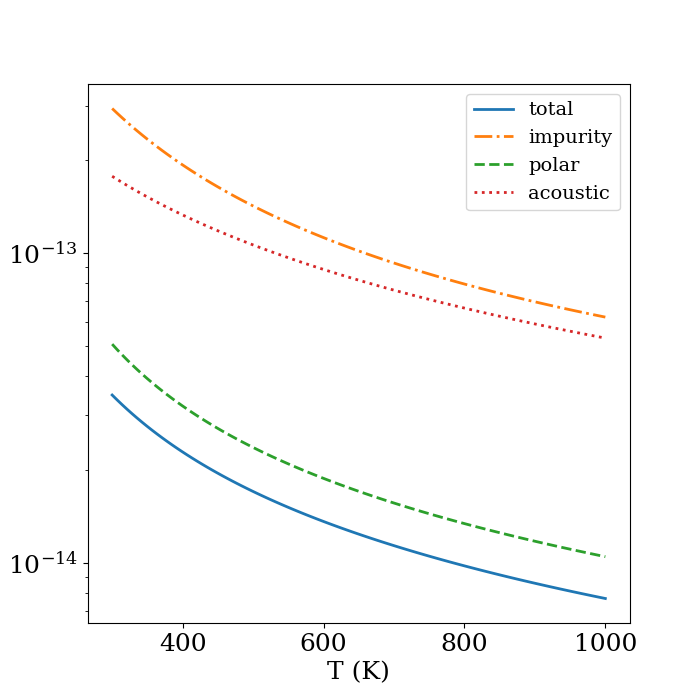}\includegraphics[width=0.5\linewidth]{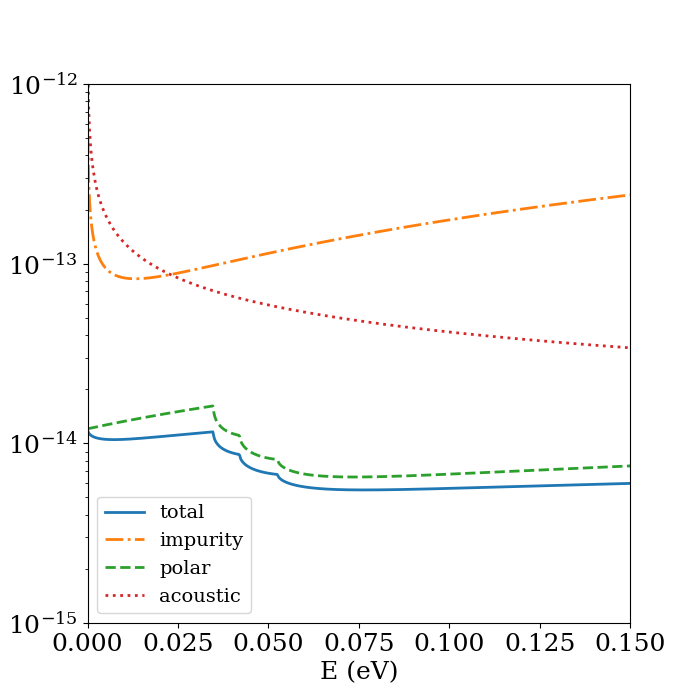}}
\caption{\label{figtau} Left: $\tau$ and its components  vs T at fixed $E$=40 meV for Mg$_3$Sb$_2$. Right: same  vs $E$  at fixed T=500 K. Optimal doping assumed in both cases.}
\end{figure} 
In Figure \ref{figtau} we display the behavior of $\tau$(T,$E$) and its components as function 
of T (left panel) at  energy fixed  at 40 meV above the band edge (the 
Fermi energy for optimal doping) and as function of energy at fixed T (we assume  optimal 
doping, discussed below, in all cases). The structure in the $E$ dependence is due to the 
three LO modes. The total $\tau$ is clearly dominated by polar scattering, due mainly to low 
LO phonon  energies, to which it is inversely proportional. Further detailed studies of the 
Fr\"olich interaction in this material remain necessary, although it is encouraging for the 
present simplified estimate that the energy average (not shown) of $\tau$(T, $E$) is in 
good qualitative agreement with mobility calculations including wavevector dependence for 
other polar materials \cite{vdw}.  

The other technical parameters in the calculations are as follows.  The energy bands to be used in Eq.\ref{eq:2.2}  are calculated on a 24$\times$24$\times$18 k-point grid in the Brillouin zone for the Mg$_3$Sb$_2$ structure optimized following quantum forces. The eigenvalues are then interpolated by a Fourier-Wannier technique \cite{bt} over a finer grid comprising a number of k-points given by the original number of points times an amplification factor, which we choose to be $A$=20 (the approximate grid being therefore 65$\times$65$\times$48).
This setting is quite sufficient to converge, for example, the Seebeck coefficient to within   1\% compared with a 12$\times$12$\times$12 ab initio grid, depending on T. 
For convenience of the computational interface, we used  
 the projector augmented wave method in the VASP code (using the maximum suggested cutoff) \cite{vasp} for the bands (they are very close to those obtained by other ab initio codes). All ab initio calculations are within density functional theory in the generalized gradient approximation \cite{pbe}.

Before turning to the results, we point out two additional issues. First, we always use the total  thermal conductivity in the ZT calculation. The electronic component, which we will discuss below, cannot be neglected in comparison to the fairly small lattice component. For the lattice component, we use our calculated   value for the polycrystal (Ref.\cite{kappa}, Fig.3) as  Mg$_3$Sb$_2$ and related alloys are generally heavily polycrystalline (of course  ZT  would be strongly diminished if we used the much larger  lattice thermal conductivity of the crystal). 
Second, although the polycrystal thermal lattice conductivity is used, we calculate the electronic transport coefficients for the perfect crystal; this is not inconsistent, at least quantitatively: within the model of Ref.\cite{gb} for the reduced conductivity due to grain boundaries, we estimated that the effects are marginal in our regime (less that 0.5-1\% on all quantitites at 10$^{19}$ cm$^{-3}$ and 300 K).

\begin{figure}[ht]
\centerline{\includegraphics[width=0.5\linewidth]{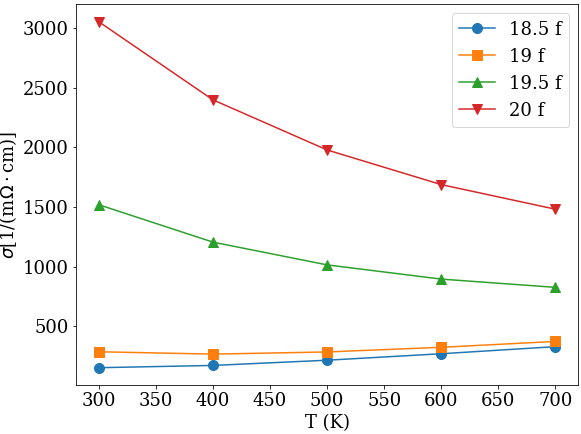}
\includegraphics[width=0.44\linewidth]{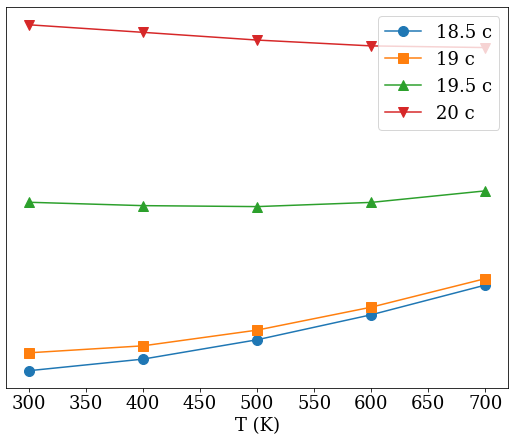}}
%\caption{\label{fig1} Electrical conductivity computed with energy-dependent (left) and constant (right) scattering time. The logarithm of doping is indicated in the labels.  The scale is intentionally the same in both Figures.}
%\end{figure} 
%\begin{figure}[ht]
\centerline{\includegraphics[width=0.5\linewidth]{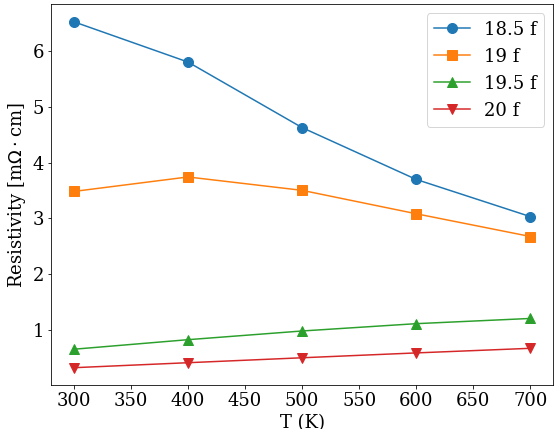}}
%\caption{\label{rho} Electrical resistivity (energy-dependent scattering time). The logarithm of doping is indicated in the labels.}
\caption{\label{fig1} Top left: electrical conductivity computed with energy-dependent scattering time. Top right: same, with constant scattering time. Bottom: electrical resistivity for energy-dependent scattering time. The logarithm of doping is indicated in the labels.}
\end{figure} 

\section{Results} 
\begin{figure}[ht]
\centerline{\includegraphics[width=0.5\linewidth]{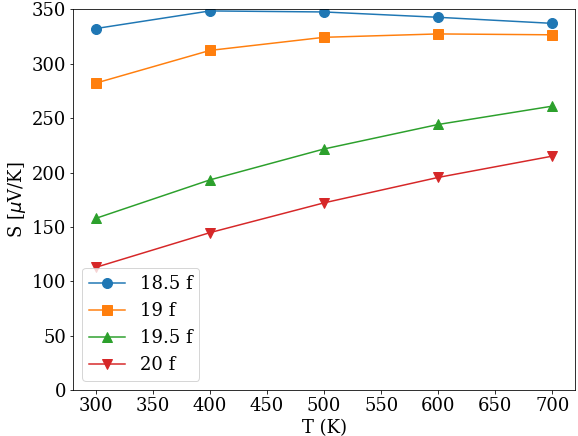}
\includegraphics[width=0.44\linewidth]{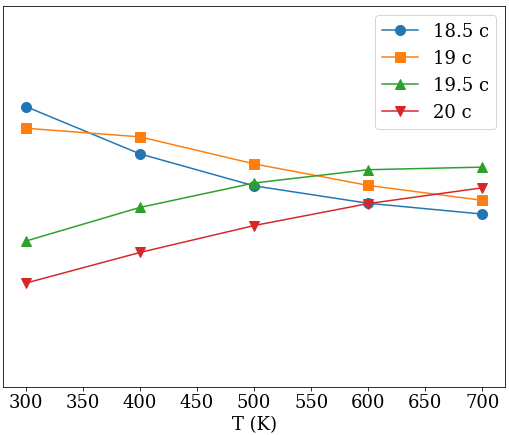}}
\caption{\label{fig2} Absolute value of the Seebeck coefficient computed with energy-dependent (left) and constant (right) scattering time. The logarithm of doping is indicated in the labels. The scale is intentionally the same in both Figures.}
\end{figure}

\begin{figure}[ht]
\centerline{\includegraphics[width=0.5\linewidth]{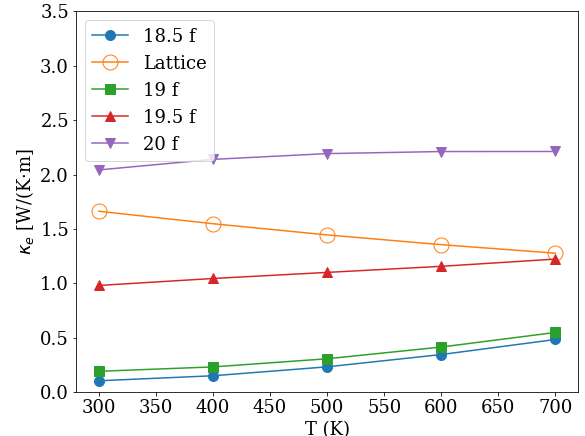}
\includegraphics[width=0.44\linewidth]{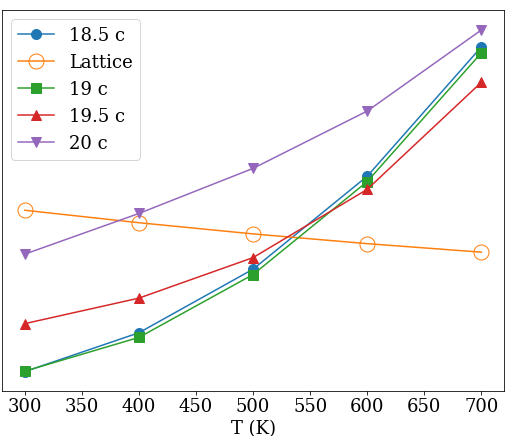}}
\caption{\label{fig3} Electronic thermal conductivity computed with energy-dependent (left) and constant (right) scattering time. The logarithm of doping is indicated in the labels.  The scale is intentionally the same in both Figures. The calculated lattice component \cite{kappa} is also displayed for reference.}
\end{figure}

In Figures \ref{fig1} to \ref{fig3}  we compare the thermoelectric coefficients calculated in the  energy-dependent relaxation time approximation (labeled `f' for `full'), in the left panel, with those in the constant-time  (labeled `c' for `constant') in the right panel. For each case, several doping levels are considered; the doping is always $n$-type.
Starting from Figure \ref{fig1} we note that the conductivity is degenerate-like only above 10$^{19}$ cm$^{-3}$, whereas below that value it recalls the activated behavior of a doped semiconductor. Empirically, therefore, this is the threshold for degenerate  behavior (observed in several cases of alloyed MgSb samples of increasing nominal doping \cite{mgsb1}). Constant scattering time produces a $\sigma$ that hardly decreases  at high T, where as in the full case phonon scattering increases in that limit, causing a drop in the conductivity. When displayed as a resistivity as in Figure \ref{fig1}, bottom, our full result is  very close to experiment (Fig.3c of Ref.\cite{mgsb1}) for the degenerate regime (the lowest curves).

The Seebeck coefficient is shown (in absolute value) in  Figure \ref{fig2}. It is consistent  with general expectations, observations, and previous calculations \cite{mgsb1} (Fig.4b and 3c), especially in the developed degenerate regime, where it is monotonically increasing with T, with values in general agreement with experiment \cite{mgsb1}. Its values in the whereabouts of 150-250 $\mu$V/K are interesting, though not exceptional: the material's  sixfold conduction band bottom rescues it  from having just a mediocre thermopower. 
At lower densities the Seebeck is larger mainly (as dictated by the Mott-Cutler formula) because of the low conductivity. The behavior in the constant time approximation is similar at high doping, although the resulting values are smaller (due to the fact that, in the full calculation, relaxation time no longer cancels out of S); at low density the Seebeck decreases with T because of the $\sigma$  overestimate due to the neglect of phonon scattering.

\begin{figure}[ht]
\centerline{\includegraphics[width=0.5\linewidth]{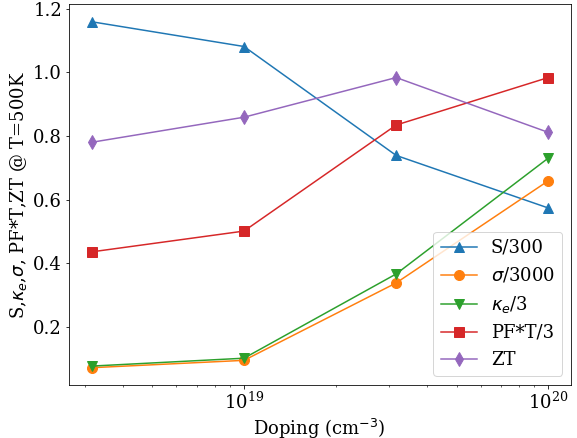}
\includegraphics[width=0.5\linewidth]{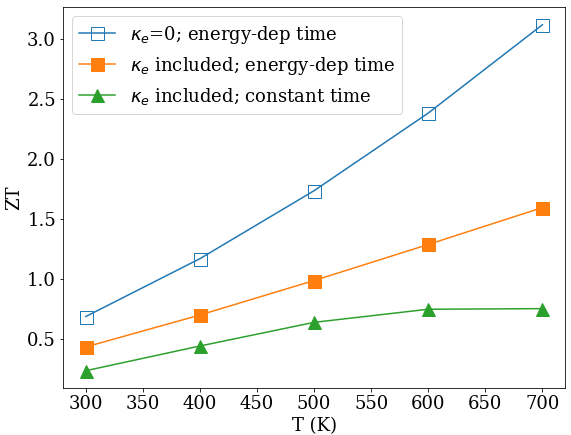}}
\caption{\label{fig6} Summary of thermoelectricity in Mg$_3$Sb$_2$. Left, doping dependence Seebeck, electronic thermal conductivity, power factor, and ZT as function of doping density (some quantities are rescaled as indicated). Right, figure of merit 
ZT at optimal doping for constant $\tau_0$ and for energy-dependent $\tau$, with electronic thermal conductivity or without it. The full ZT to make reference to is the filled-squares curve.}
\end{figure}

Figure \ref{fig3} shows that the electronic thermal conductivity $\kappa_e$ is significant on the scale set by  the lattice component $\kappa_{\ell}$ of  the polycrystal (also shown for reference, as calculated by us in Ref.\cite{kappa}); of course,  the much larger  lattice thermal conductivity of the crystal would dominate over $\kappa_e$. 
For all T at the  optimal doping (labeled `19.5 f' in the Figure, and discussed below in reference to Figure \ref{fig6}), $\kappa_e$ and $\kappa_{\ell}$ are similar. Not only is it   important to account for $\kappa_e$ in the determination of ZT (see Figure \ref{fig6} below), but also the different descriptions of scattering lead to qualitatively different predictions. In the constant-time approximation, $\kappa_e$ becomes larger than in the energy-dependent case and rises sharply at high temperature, again due to the neglect of phonon scattering. This causes a strong increase of the total thermal conductivity at high T. In the energy-dependent case (in particular at optimal doping), instead, the total thermal conductivity will still decrease slightly with temperature, which is usually the case in experiment \cite{mgsb1}.

In Fig.\ref{fig6}, left, we collect all coefficients  as function of doping, choosing arbitrarily T=500 K. As expected, Seebeck decreases, and $\sigma$ increases. In this range of doping, the power factor P=$\sigma S^2$ has not yet peaked, though it seems likely  to do so around 10$^{20}$ cm$^{-3}$.  Due to the sharp raise of $\kappa_e$, tracking that of the electrical conductivity, ZT does instead peak at about 3$\times$10$^{19}$ cm$^{-3}$, which is then by definition the optimal doping. This value is practically unaffected by temperature in the range we consider. We note that, because of the very definition of ZT, the lattice conductivity $\kappa_L$ is crucial in determining the optimal doping and its T dependence; in particular, ZT would have no maximum (monotonic increase at all T)  in our density range if we used the crystal $\kappa_L$; or it would have different maxima at different T's if we used intermediate values between poly and crystal.

Finally, we show the figure of merit ZT for our  $n$-doped  polycrystal in Figure \ref{fig6}, right. To reduce clutter, we pick the optimal doping and compare the full ZT in  the energy- and temperature-dependent relaxation-time approximation (filled squares) with that in the constant-time approximation (triangles). The full ZT tops at 1.6 at the top of our T range, and the T-trend and values  are in good agreement with experiment for the same kind of material (see Figure 1a of Zhang {\it et al.}, Ref.\cite{mgsb1}).
The constant-$\tau$ case is about a factor of 2 smaller, and has a clear maximum in T, mainly due to the larger increase in $\kappa_e$ with T in that case. Besides the two main approximations, in Figure \ref{fig6} we also show ZT calculated neglecting the electronic thermal conductivity (empty squares). Clearly this neglect artificially  enhances  ZT by  over a factor 2. This explains the  overly optimistic values of  2.5 to 3 reported in recent work \cite{chino}, where  electronic thermal conductivity was not included.

\section{Summary and acknowledgments}

We have calculated the thermoelectric coefficients and figure of merit ZT in Mg$_3$Sb$_2$ using a temperature- and  energy-dependent scattering time. The effects of this improved approximation are significant, especially for the electrical and thermal conductivities. The ab initio calculated lattice thermal conductivity is included.  Where applicable, our results are in good agreement with experiment. 
The thermopower is good, though not exceptional, and  the small lattice thermal conductivity is in fact a major co-factor in producing an interesting figure of merit. ZT increases monotonically in the T range we consider; on the other hand, it has a maximum as a function of doping: our estimate of the peak-ZT optimal $n$-doping of Mg$_3$Sb$_2$ is about 3$\times$10$^{19}$ cm$^{-3}$, over all the 300 to 700 K range.
 
Work supported in part by UniCA, Fondazione di Sardegna, and Regione Autonoma della Sardegna via Progetto biennali di ateneo 2016 {\it Multiphysics approach to thermoelectricity}, by CINECA, Bologna, Italy, through ISCRA Computing Grants, and by CRS4 Computing Center, Piscina Manna, Italy. 

\end{document}